\documentclass[12pt,]{article}
 \usepackage{graphicx}
 \usepackage[cp1251]{inputenc}
 \usepackage{epstopdf}
 \usepackage{rotating}
\begin{document}

\noindent{\it Publications of the Pulkovo Observatory (ISSN 0367-7966), Issue 227}

 \bigskip
 \bigskip

\centerline{\large\bf Study of the structure and kinematics of the Galaxy according to}
\centerline{\large\bf  VLBI-astrometry of masers and radio stars}
 \bigskip
 \centerline{
            V.V.~Bobylev \footnote[1]{{\it e-mail:} vbobylev@gaoran.ru},
            A.T.~Bajkova  \footnote[2]{{\it e-mail:} bajkova@gaoran.ru}
            }
 \bigskip
 \centerline{\small\it Central (Pulkovo) Astronomical Observatory, RAS, 65/1, Pulkovskoye Ch., 196140, St.-Petersburg}
 \bigskip
 \bigskip

 %\begin{abstract}
In recent years, radio interferometric observations have achieved high accuracy in determining the absolute values of trigonometric parallaxes and proper motions of maser radiation sources and radio stars. The error in determining the trigonometric parallaxes of these objects averages about 10 microarcseconds, which allows us to confidently study the geometric and kinematic properties of the distribution of stars located at great distances from the Sun, up to the center of the Galaxy. This article provides an overview of the main results of studying the structure and kinematics of the Galaxy, which were obtained by various scientific teams using VLBI observations of masers and radio stars. The main attention is paid to the results of studying the Galaxy obtained by the authors of this work.

  \bigskip
  \bigskip

  \noindent
 {\it Keywords:} Masers: Galaxy (Milky Way): structure and kinematics

\newpage
\section*{Introduction}
Masers are emitted in the immediate vicinity of young forming stars, as well as already evolved stars pumped either by strong infrared radiation or by strong gas collisions in disks, jet streams or winds. A feature of masers is that almost all of their energy is emitted in several molecular lines. These are, for example, hydroxyl (OH) masers with a frequency of 1.6~GHz, methanol masers (CH$_3$OH) with a frequency of 6.7 and 12.2~GHz, water vapor masers (H$_2$O) with a frequency of 22~GHz or silicon monoxide (SiO) masers with a frequency of 43 GHz.

The sources of maser radiation associated with very young stars and protostars located in regions of active star formation are of great interest for studying the structure and kinematics of the Galaxy. First of all, this is due to the fact that recently it has been possible to measure trigonometric parallaxes and proper motions of such objects with high accuracy. For this, observations are carried out using very long baseline radio interferometers (VLBI). The value of radio observations lies in the fact that they are not hindered by the absorption of radiation by interstellar dust.

Since 2000, regular VLBI observations of masers in regions of active star formation began in order to determine their high-precision absolute trigonometric parallaxes and proper motions. In addition to the main source, several reference extragalactic objects (quasars) located no further than $\sim2.5^\circ$ from the target object were observed. To reliably obtain a parallactic ellipse, long-term observations are required, preferably dense ones. However, maser spots are not stable. It happens that in a short time some spots in the observed source fade, while others suddenly appear.

Internal motions in the envelopes of host stars reach velocities of several tens of km s$^{-1}$. Therefore, to determine reliable average source velocities, it is necessary to track as many maser spots as possible.

Nevertheless, in 2006 (Xu et al., 2006; Hachisuka et al., 2006) the first results of determination of high-precision absolute parallaxes and proper motions of maser sources in the star-forming region W3\,(OH), which is located in the Perseus arm, appeared. To date, there are about 200 such definitions for various regions in the Galaxy (Reid et al., 2019; Hirota et al., 2020).

Masers associated with massive young stars have a huge luminosity, so they can be observed in the distant regions of the Galaxy. Moreover, megamasers observed in other galaxies are known (Gao et al., 2017). To study the structure and kinematics of the Galaxy, the results of VLBI observations of young radio stars in the continuum are also of interest. These are mainly low-mass stars of the T\, Tauri type, located in the Local Arm. There are also high-precision VLBI observations in the continuum of massive high-luminosity binary stars that are black hole companions, for example, the star V404\,Cyg (Miller-Jones et al., 2009) or GRS\,1915$+$105 (Reid et al., 2014 ).

As a result, more than 250 VLBI measurements of absolute parallaxes and proper motions of masers and radio stars have been published so far. For all these objects, the values of their systemic radial velocities are known, which are mainly measured not from maser, relatively wide spectral lines, but from narrower emission lines of gas clouds surrounding the source (from CO lines).

With the accumulation of measuring material, works periodically appear devoted to the analysis of the structure and kinematics of the Galaxy using data on masers in regions of active star formation. Noteworthy are the works of Reid et al. (2009), Bobylev, Bajkova (2010), Honma et al. (2012), Rastorguev et al. (2017), Reid et al. (2019), Hirota et al. (2020), Bobylev, Bajkova (2022). The work of Immer and Rygl (2022) is also of great interest, where a detailed description of the parameters of the spiral structure obtained from masers is given.

This paper reviews the main results of studying the structure and kinematics of the Galaxy, which were obtained by various research teams using VLBI observations of masers and radio stars, in particular, the authors of this article. These are such parameters as the angular velocity of rotation of the Galaxy $\Omega_0$ and its derivatives $\Omega'_0,$ $\Omega''_0$, the distance from the Sun to the center of the Galaxy $R_0$, the geometric and kinematic parameters of the galactic spiral density wave, as well as the amplitudes of perturbations of the vertical positions and vertical velocities of the stars in the thin disk of the Galaxy.

\section{VLBI observations}
The results of VLBI observations of masers associated with young stars and protostars are combined in the BeSSeL survey project (The Bar and Spiral Structure Legacy Survey \footnote[1]{http://bessel.vlbi-astrometry.org}). The project is aimed at determining high-precision distances to star-forming regions, studying the structure, kinematics and dynamics of the Galaxy. The most important contributor here is the US VLBA array, which consists of ten 25-meter antennas with a maximum baseline of over 8000~km. The observations cover frequencies of 6.7 and 12.2~GHz with methanol maser transitions, as well as water vapor maser transitions at a frequency of 22.2~GHz.

Another contributor to the BeSSeL review is the European VLBI network EVN (European VLBI Network). Here, the longest baselines are about 9000~km, and the largest in the array is the 100~m antenna at Effelsberg. Observations are carried out at frequencies from 6.7 to 22.2 GHz.

In Japan, VLBI observations of masers are being carried out using the VERA program (VLBI Exploration of Radio Astrometry \footnote[2]{http://veraserver.mtk.nao.ac.jp}). The interferometer consists of four 20-meter antennas located throughout Japan, providing a base length of 1020 to 2270~km. Observations are being made of H$_2$O masers at a frequency of 22.2~GHz, less often, SiO masers at a frequency of 43.1 and 42.8~GHz. The most important unique feature of VERA antennas is a two-beam receiving system that allows simultaneous tracking of a pair of maser targets and phase reference sources. In all other programs (VLBA, EVN, etc.), observations of reference extragalactic objects are made at the beginning and end of the session by redirecting antennas, which then requires additional efforts to take atmospheric distortions into account. Note that the astrometric accuracy is the better, the higher the frequency of observations. Thus, the VLBI observations made under the VERA program are the most accurate in comparison with the observations obtained within the framework of other programs.

The East Asian VLBI network, EAVN (East Asian VLBI network
\footnote[3]{https://radio.kasi.re.kr/eavn/main$\_$eavn.php }). Here, the contributors are the Korean KVN (Korean VLBI Network), the Chinese CVN (Chinese VLBI Network) and the Japanese, VERA, VLBI networks. At present, EAVN consists of 21 telescopes that are used to observe H$_2$O masers at a frequency of 22.2~GHz. The results of observations of the source G~034.84$-$00.95 are reflected in the work of Sakai et al. (2022).

The first result of the VLBI measurement of the source parallax G~339.884$-$1.259, obtained with the LBA (Long Baseline Array) radio interferometer in Australia, is also known (Krishnan et al., 2015). The interferometer consisted of five antennas of large diameter (more than 20 meters), methanol masers were observed at a frequency of 6.7~GHz.

In addition to masers, VLBI observations of young radio stars in the continuum are of interest. Currently, about 60~stars observed by the GOBELINS program (Ortiz-Le\'on et al., 2017) at frequencies of 5 and 8~GHz, there is complete information~--- their absolute trigonometric parallaxes and proper motions have been measured, and their line-of-sight velocities are also known.

\begin{figure}[t]
    \centering
    \includegraphics[width=0.95\textwidth]{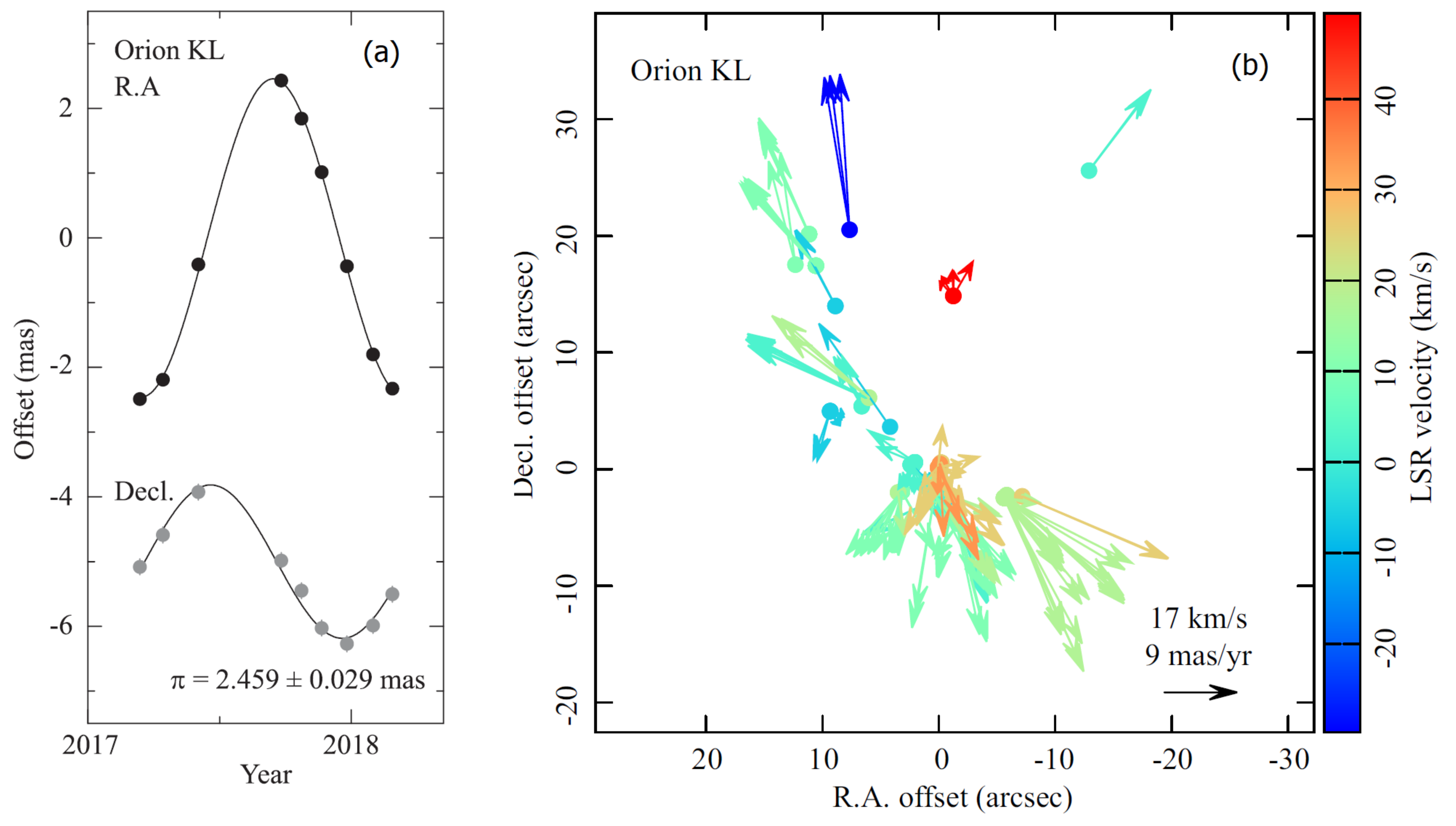}
    \caption{
An example of VLBI observation using the VERA program for H$_2$O-masers in the Orion\,BN/KL region. Periodic changes in $\alpha$ and $\delta$ coordinates after deletion of trends describing proper motions of the source center~(a). Vectors of relative proper motions of individual maser spots around the source~(b). The line-of-sight velocities are given relative to the Local Standard of Rest (LSR). Figure is taken from Nagayama et al. (2020).
}
    \label{f-Ori-KL}
\end{figure}
%%%%%%%%%%%%%%%%%%%%%%%%%%%%%%%%%%%%%%%%%
\begin{figure}[t]
    \centering
    \includegraphics[width=0.95\textwidth]{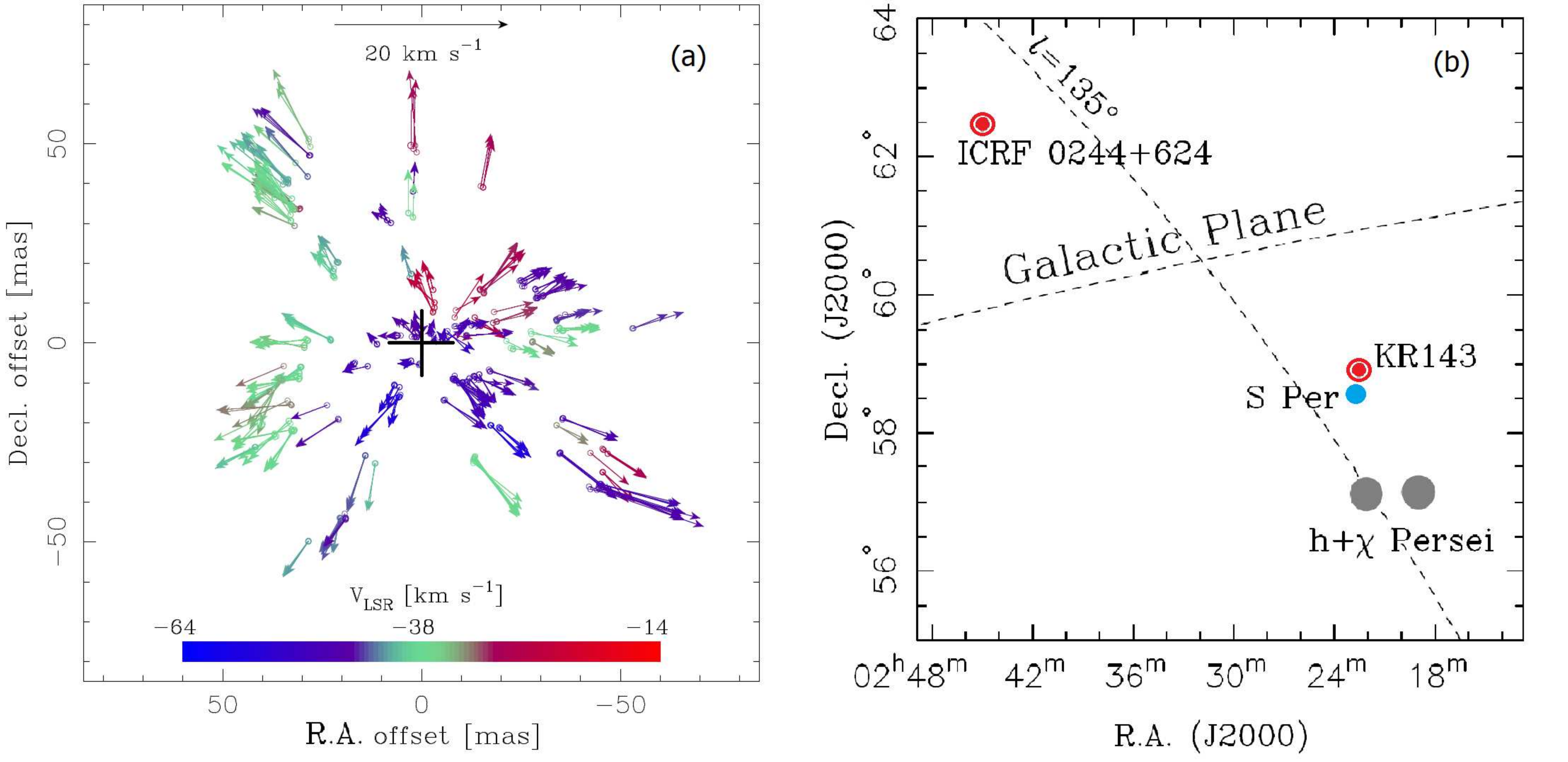}
    \caption{
An example of VLBI observations of H$_2$O masers in the envelope of a red supergiant S\,Per with a VLBA array. Vectors of relative proper motions of individual maser spots around the source~(a), position S\,Per on the celestial sphere, near the extragalactic radio source KR\,143~(b). Figure is taken from Asaki et al. (2010).
}
    \label{f-S-Per}
\end{figure}

Figure~\ref{f-Ori-KL} shows an example of VLBI observations of H$_2$O masers associated with a radio source in the Orion\,BN/KL region (Asaki et al., 2010) using the VERA program. We see that the distribution of individual maser spots in the source region is by no means symmetrical. The velocity vectors in Fig.~\ref{f-Ori-KL}(b) reflect the internal kinematics of the object.

To determine the trigonometric parallax of the target source, 1.5--2 years of regular observations are required. As already noted, the first results of VLBI observations of masers were published in 2006. Already in 2009, enough such measurements were accumulated to be used to estimate the parameters of the rotation of the Galaxy (Reid et al., 2009). At present, the absolute (with reference to extragalactic reference sources) trigonometric parallaxes and proper motions of more than 200 masers have already been measured by the VLBI method. The achieved error in determining trigonometric parallaxes is on average about 10 microarcseconds. This makes it possible to confidently (with an error of less than 10\%) analyze objects located from the Sun up to 10~kpc, i.e., up to the center of the Galaxy.

Figure~\ref{f-S-Per} shows the results of VLBI observations using the US VLBA array at the NRAO (National Radio Astronomy Observatory) radio observatory of masers in the shell of the red supergiant S\,Per (Asaki, et al., 2010). It can be seen in Fig.~\ref{f-S-Per}(a) that more than 40 maser spots are very evenly distributed in a region with a radius of about 50~mas (milliseconds of arc) around the center of the observable envelope, and the residual velocity vectors perfectly indicate the position of the center, which is marked with a cross. Figure~\ref{f-S-Per}(b) shows the position of the well-known double open star cluster $h$ and $\chi$~Perseus, whose members are very young OB stars. And most importantly~--- very close to the star S\,Per there is an extragalactic radio source KR\,143, binding to which serves to determine the absolute values of trigonometric parallaxes and proper motions. There is one more extragalactic radio source in the field of view~--- ICRF\,0244$+$624, quite far from S\,Per, but in Asaki, et al., (2010) it was also used to take into account the tropospheric delay and control stability of the source KR\,143.

The H$_2$O masers around S\,Per were monitored by Asaki\,et al., (2010) for six years. VLBI observations were carried out at a frequency of 22~GHz. As a result, a first-class result was obtained, $\pi=0.413\pm0.017$~mas, $\mu_\alpha\cos\delta=-0.49\pm0.23$~mas/year and $\mu_\delta=- 1.19\pm0.20$~mas/year. Thus, the estimate of the heliocentric distance to S\,Per was $2.42^{+0.11}_{-0.09}$~kpc, and a very small deviation from the circular velocity of rotation around the center of the Galaxy was found, about 12~km s$^{-1}$ (is the peculiar velocity of the star).

\begin{figure}[t]
    \centering
    \includegraphics[width=0.95\textwidth]{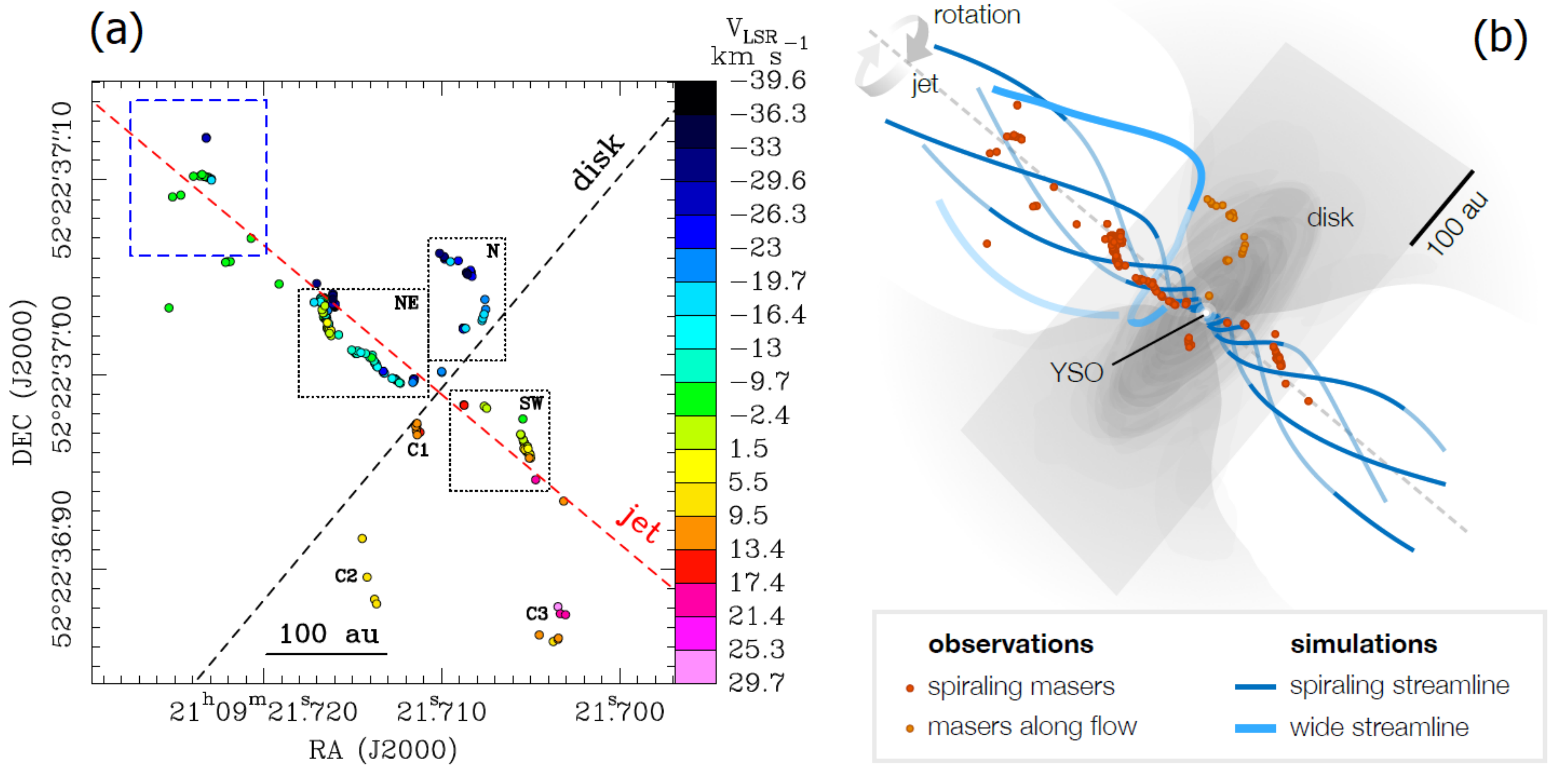}
    \caption{
Distribution on the celestial sphere of H$_2$O-masers associated with the radio source IRAS 21078$+$5211~(a), the position of these maser spots along with the proposed disk and jet model~(b). Figure is taken from Moscadelli et al. (2022).
}
    \label{f-Disk-Jet}
\end{figure}

The most complex and confusing observed picture is given by masers associated with forming protostars and the youngest stars. Here (Figure~\ref{f-Disk-Jet}), in one region, there are masers associated with a disk rotating around the star, masers in a jet, masers in the environment, etc. with a very young forming massive ($5.6\pm2~M_\odot$) star IRAS 21078$+$5211, which has a disk and a magnetized jet. In the presence of a sufficient number of maser spots with measured line-of-sight velocities and with a favorable source orientation, the observed pattern can be deciphered, which was done in the recent work of Moscadelli et al. (2022).

So far, there is one drawback~--- almost all observations of masers and radio stars are made from the northern hemisphere of the Earth. Therefore, in the projection onto the Galactic plane, the fourth galactic quadrant remains uncovered by sources.

\begin{figure}[t]
    \centering
    \includegraphics[width=0.75\textwidth]{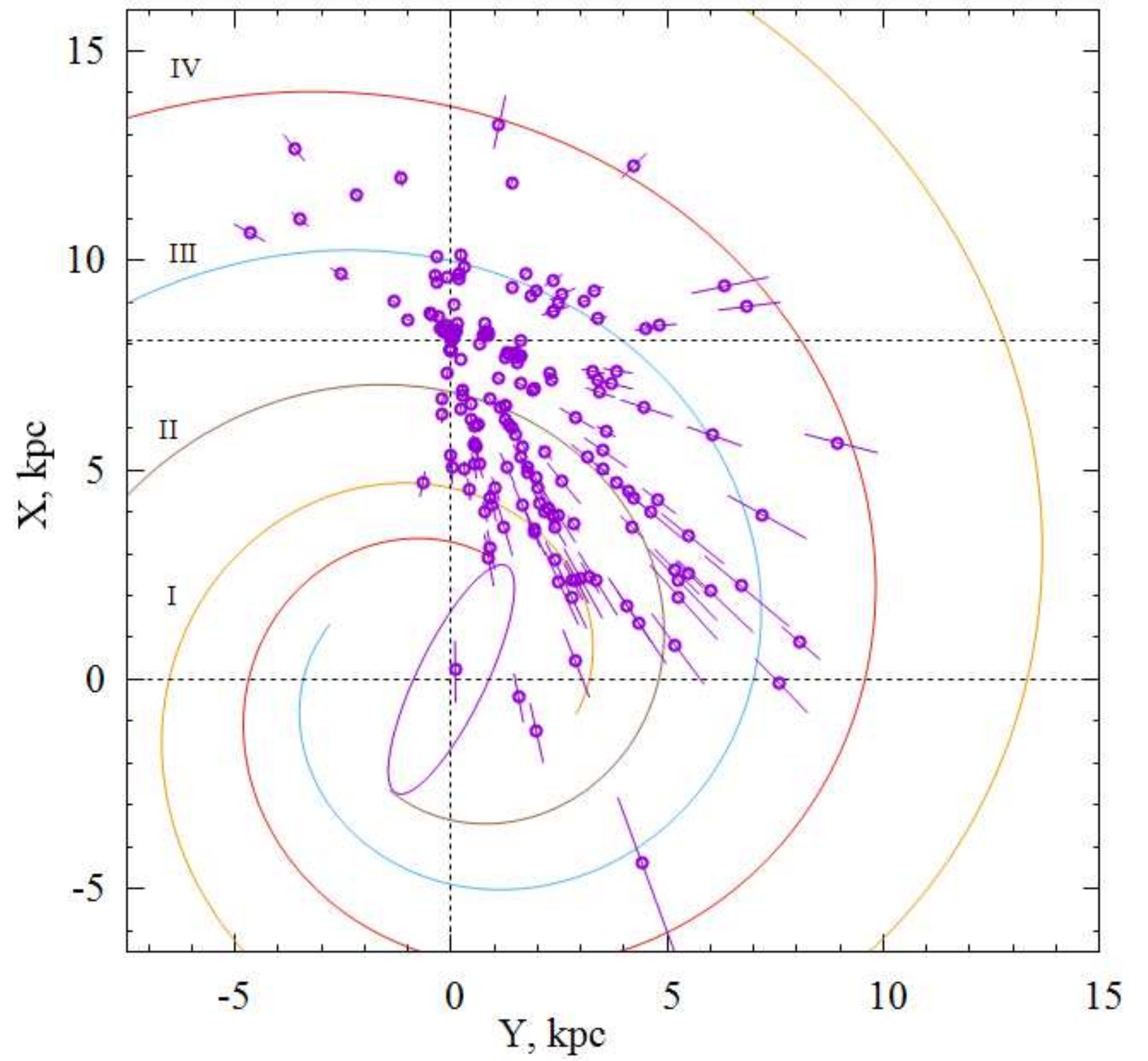}
    \caption{Distribution of masers and radio stars with trigonometric parallax errors less than 15\% in the projection on the galactic plane $XY$, a four-armed spiral pattern with a pitch angle of $i=-13^\circ$ is shown, the central galactic bar is marked, the Sun is located at the point $(X,Y)=(8.1,0)$~kpc.}
    \label{f-15-XY}
\end{figure}

Figure~\ref{f-15-XY} shows an example of the distribution of masers and radio stars projected onto the galactic plane $XY$. To construct the figure, a sample of 256 objects was used, from which 204 masers with relative parallax errors less than 15\% were selected. A coordinate system is used in which the $X$ axis is directed from the center of the Galaxy to the Sun, the direction of the $Y$ axis coincides with the direction of the Galaxy's rotation. A four-armed spiral pattern with a pitch angle $i=-13^\circ$ is given according to Bobylev and Bajkova (2014), here it is plotted with the value of the distance from the Sun to the axis of rotation of the Galaxy $R_0=8.1$~kpc, the following four are numbered in Roman numerals spiral arms: I~--- Scutum, II~--- Carina-Sagittarius, III~--- Perseus and IV~--- the Outer arm.

%%%%%%%%%%%%%%%%%%%%%%%%%%%%%%%%%%%%%

 \section{Structure and kinematics of the Galaxy}
For studying the dynamics of the Galaxy, the nature of the rotation curve is of paramount importance. On its basis, the mass of the Galaxy, the nature of the distribution of matter density in all directions in the Galaxy are estimated, the kinematic properties of various galactic subsystems, etc. are studied. In particular, masers associated with young stars and protostars are of particular interest. They are representatives of the extremely thin disk of the Galaxy; they rotate as quickly as possible around the axis of rotation of the Galaxy.

To plot the rotation curve of the Galaxy, such quantities as the distance from the Sun to the center of the Galaxy $R_0$ (more precisely, from the Sun to the axis of rotation of the Galaxy), the angular and linear rotation velocities at the near-solar distance $\Omega_0$ and $V_0$ are important, while $V_0= R_0\Omega_0$. Various approaches are used to estimate the values of these parameters. Sometimes the linear rotation velocity and its derivatives are directly estimated from observational data (Reid et al., 2009; Hirota et al., 2020). The method for estimating the angular velocity of rotation and its derivatives is more common (Rastorguev et al., 2017; Bobylev et al., 2022).

%%%%%%%%%%%%%%%%%%%%%%%%%%%%%%%%%%%%%%%%%%%%%%%%%%%%%%%%%%%%%% t-R0
 \begin{table}[t] \caption[]{\small
Estimates of distance $R_0,$ velocities $\Omega_0,$ $\Omega'_0,$ $\Omega''_0$ and $V_0$ found from masers and radio stars. }
  \begin{center}  \label{t:R0}
  \small\begin{tabular}{|c|c|c|c|c|c|c|}\hline
  $R_0$ & $n_\star$ & $\Omega_0$ & $\Omega'_0$ & $\Omega''_0$ & $V_0$ & Ref \\
 kpc & & km/s/kpc & km/s/kpc$^2$& km/s/kpc$^3$ & km/s & \\\hline
 $8.03\pm0.12$&  93& $ 29.7\pm0.5 $& $-4.20\pm0.11$ & $0.73\pm0.03$ & $238\pm6$ & (1)\\
 $8.19\pm0.12$& 131& $28.64\pm0.53$& $-4.00\pm0.09$ & $1.28\pm0.04$ & $235\pm7$ & (2)\\
 $8.15\pm0.15$& 199& $28.95\pm0.27$&                &               & $236\pm7$ & (3)\\
 $7.92\pm0.16$&  99& $28.63\pm0.26$&                &               & $227\pm5$ & (4)\\
 $8.15\pm0.12$& 256& $29.01\pm0.33$& $-3.90\pm0.07$ & $0.83\pm0.03$ & $236\pm4$ & (5)\\
 $8.1\pm0.1$~*& 150& $30.18\pm0.38$& $-4.37\pm0.08$ & $0.85\pm0.04$ & $244\pm4$ & (6)\\
  \hline
  \end{tabular}\end{center}
 {\scriptsize
 $n_\star$~is a number of masers in the sample,
 (1)~-- Bajkova, Bobylev (2015),
 (2)~-- Rastorguev et al. (2017),
 (3)~-- Reid et al. (2019),
 (4)~-- Hirota et al. (2020),
 (5)~-- Bobylev et al. (2020),
 (6)~-- Bobylev, Bajkova (2022),
 (*)~-- $R_0$ was taken as predetermined according to the recommendation of Bobylev and Bajkova (2021).
 }
 \end{table}
%%%%%%%%%%%%%%%%%%%%%%%%%%%%%%%%%%%%%% t-R0

 \subsection{Distance from the Sun to the center of the Galaxy}
The table~\ref{t:R0} gives estimates of the distance $R_0$, as well as the velocities
$\Omega_0,$ $\Omega'_0,$ $\Omega''_0$ and $V_0$ found from different samples of masers and radio stars. Here we mean the following expansion of the angular velocity of rotation $\Omega$ into a series up to terms of the corresponding order of smallness $r/R_0:$
\begin{equation} \begin{array}{lll}
 \Omega=\Omega_0+\Omega^{\prime}_0(R-R_0)+\Omega^{\prime\prime}_0(R-R_0)^2+...,
 \label{EQ-1}
 \end{array} \end{equation}
where $r$~is the heliocentric distance of the star, which is calculated through the trigonometric parallax $r=1/\pi$, $R$~is the distance from the star to the axis of rotation of the Galaxy $R^2=r^2\cos ^2 b-2R_0 r\cos b\cos l+R^2_0$.
Sometimes the expansion of the angular velocity of rotation $\Omega$ by a polynomial in inverse powers of $R$ is used (Loktin, Popova, 2019).

Almost all the values indicated in the table~\ref{t:R0} were found simultaneously, i.e., $R_0$ also acted as the desired unknown when solving the kinematic equations. True, in each of these works, the equations were of a slightly different form. Masers located in the inner region of the Galaxy ($R<4$~kpc), where there is a noticeable influence of the central bar, were not considered.

The results of the determination of $R_0$ (the first five rows of the table~\ref{t:R0}) can be compared with the most reliable modern estimate obtained from the analysis of a sixteen-year series of observations of the motion of the star S2 around the massive black hole Sgr~A* in the center of the Galaxy $R_0= 8.178\pm0.013$~(stat.)$\pm$0.022~(sys.) kpc (Abuter et al., 2019). However, the latest publication of this team (Abuter et al., 2021) shows the presence of instrumental aberrations. Therefore, all previous estimates of the collaboration, starting from 2018, were revised, and an updated value $R_0=8.275\pm0.009$~(stat.)$\pm$0.033~(sys.) kpc was proposed.

Note that in Bobylev and Bajkova (2021), the value $R_0=8.15\pm0.11$~kpc was derived as a weighted average from a large number of modern individual estimates that were obtained from 2011 to 2021. And the value $R_0=8.1\pm0.1$~kpc was recommended for practical use. On the whole, we can conclude that the estimates of $R_0,$ found from masers are in very good agreement with the most reliable modern estimates of this quantity.

\begin{figure}[t]
    \centering
    \includegraphics[width = 0.85\textwidth]{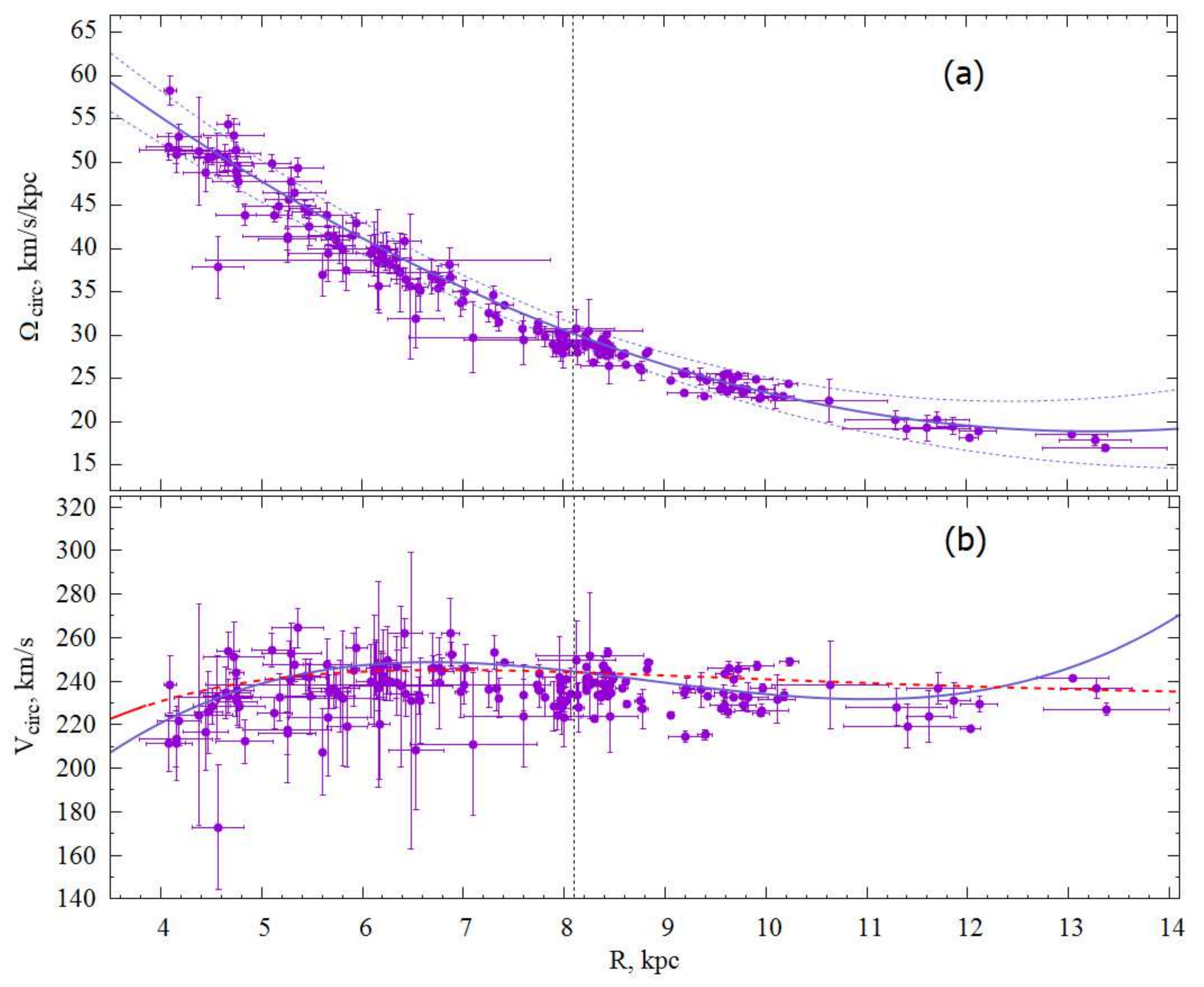}
    \caption{
Angular rotation velocities $\Omega_{circ}$ (a) and linear rotation velocities $V_{circ}$~(b) of masers depending on the distance $R$, the vertical line marks the position of the Sun, see also the text.}
    \label{f-15-rotation}
\end{figure}

 \subsection{Galaxy rotation parameters}
The velocities $\Omega_0,$ $\Omega'_0,$ $\Omega''_0$ and $V_0$ indicated in the table~\ref{t:R0} were found at different times from samples of masers and radio stars.
The masers and radio stars analyzed in these papers are representatives of the youngest stellar population in the Galaxy. We see that they revolve around the galactic center at a fairly high speed.

Figure~\ref{f-15-rotation} shows the rotation curves of the Galaxy constructed both for the angular velocities of rotation $\Omega_{circ}$ fig.~\ref{f-15-rotation}(a) and for linear rotation velocities $V_{circ}$~fig.~\ref{f-15-rotation}(b) of masers and radio stars. The solid line in both graphs shows the rotation curves found by Bobylev and Bajkova (2022) from a sample of masers with parallax errors less than 10\%. The dotted lines in Fig.~\ref{f-15-rotation}(a) indicate the boundaries of the confidence regions corresponding to the error level 1$\sigma$.

The red dotted line in Fig.~\ref{f-15-rotation}(b) shows the rotation curve found by Reid et al. (2019), where it is given in tabular form. To construct it, Reid et al. (2019) used an approach they call the construction of a two-parameter universal rotation curve. This approach is used to study the rotation of spiral galaxies (Persic et al., 1996). When plotting Fig.~\ref{f-15-rotation}, we added 8~km/s to the data from Reid et al. (2019) to match our curve at $R_0$ (see Table~\ref{t:R0}). On the whole, we can see good agreement in the behavior of the two rotation curves over the distance interval $R$ under consideration.

 \subsection{Vertical distribution of matter}
Bobylev and Bajkova (2016b) used data on HII zones, molecular clouds, and methanol masers to estimate the parameters of the vertical distribution in the Galaxy of Young Thin Disk Objects. For this, large samples with estimates of kinematic distances were used. It was shown that the objects of the Local Arm significantly affect the nature of the vertical distribution. Deviations from the symmetry of the distribution are especially noticeable in the sample of masers with measured trigonometric parallaxes, where the proportion of masers belonging to the Local Arm is large. As a result, the objects located inside the Local Arm were excluded from consideration.

In the case of an exponential density distribution, the observed frequency distribution of objects along the $Z$ coordinate axis is described by an expression of the following form:
 \begin{equation}
  N(Z)=N\exp \biggl(-{|Z-Z_\odot|\over h} \biggr),
 \label{expon}
 \end{equation}
where $N$~is a normalization factor.

In the model of a self-gravitating isothermal disk, the observed frequency distribution of objects along the $Z$ axis is described by the following formula (Spitzer, 1942):
 \begin{equation}
  N(Z)=N{\hbox { sech}}^2 \biggl({Z-Z_\odot\over \sqrt2~h}\biggr).
 \label{self-grav}
 \end{equation}
When comparing the obtained results, it should be taken into account that in the model~(\ref{self-grav}) some authors use different coefficients in the denominator, or two
$N(Z)=N{\hbox{ sech}}^2[(Z-Z_\odot)/2h]$ (Maiz-Apell\'aniz, 2001), or one
$N(Z)=N{\hbox{ sech}}^2[(Z-Z_\odot)/h]$ (Marshall et al., 2006).

Based on the model~(\ref{self-grav}), for objects located inside the solar circle ($R\leq R_0$), estimates were obtained for the elevation of the Sun above the disk symmetry plane $Z_\odot$ and the value of the vertical scale of the disk $h$.
The following scores were obtained:
from a sample of 639 methanol masers $Z_\odot=5.7\pm0.5$~pc and $h=24.1\pm0.9$~pc,
for 878 HII zones $Z_\odot=7.6\pm0.4$~pc and $h=28.6\pm0.5$~pc,
for 538 giant molecular clouds $Z_\odot=10.1\pm0.5$~pc and $h=28.2\pm0.6$~pc.

In the case of an exponential distribution of the density~(\ref{expon}) over masers, the value $h=26.5\pm0.7$~pc was found. This value is slightly larger than $h=19\pm2$~pc found from a sample of 199 masers with measured trigonometric parallaxes by Reid et al. (2019). In addition, Reid et al. (2019) found the value $Z_\odot=5.5\pm5.8$~pc, which is consistent with our more precise estimate $Z_\odot=5.7\pm0.5$~pc.

Estimates of $h$ show that masers are representatives of an extremely thin disk population. For example, the value of the height scale obtained on the basis of the~(\ref{expon}) model for a huge sample of classical Cepheids (with an average age of $\sim100$~million years) located inside the solar circle is $h=73.5\pm3.2$~ pc (Skowron, 2019) significantly exceeds that found for masers.

\section{Spiral Density Wave}
When analyzing a sample of 28 masers, Bobylev and Bajkova (2010) showed for the first time that deviations of maser velocities from the rotation curve of the Galaxy can be explained by the influence of the galactic spiral density wave. Based on the analysis of this sample, Stepanishchev and Bobylev (2011) found that the peculiar velocity of the Sun relative to the Local Standard of Rest significantly depends on the value of the Sun's phase in the spiral density wave.

 \subsection{Galaxy spiral pattern}
The position of a star in a logarithmic spiral wave can be described by the following equation:
 \begin{equation}
 R=R_0 e^{(\theta-\theta_0)\tan i},
 \label{spiral-1}
 \end{equation}
where $\theta$~is a position angle of the star: $\tan\theta=y/(R_0-x)$, where $x,y$~are heliocentric galactic rectangular coordinates of the star, with the $x$ axis directed from of the Sun to the galactic center, and the direction of the $y$ axis coincides with the direction of the galactic rotation; $\theta_0$~is some arbitrarily chosen initial angle; $i$~a pitch angle of the spiral pattern ($i<0$ for a twisting spiral).

Since $\theta_0$ is a constant, and the approximate value of $\tan i$ is known to us from previous studies, we can take $\theta_0\tan i={\rm const}$ as a first approximation. Now the equation~(\ref{spiral-1}) after taking the logarithm can be rewritten like this:
$\ln (R/R_0)=\theta\tan i+{\rm const}$, or more conveniently:
\begin{equation}
  \ln (R/R_0)=a\theta+b.
 \label{spiral-03}
\end{equation}
As you can see, the relation~(\ref{spiral-03}) is the equation of a straight line in the plane ``position angle~--logarithm of distance''. Solving the system of conditional equations separately for each segment of the spiral arm using the least squares method, we can find two quantities: $a$ and $b$. It is obvious that $a=\tan i$. Now suppose that $\theta=0,$ then we find the value $a_0=R_0e^b$~--- the place where the center of the considered spiral arm intersects the axis $X,$ directed from the center of the Galaxy and passing through the Sun. That is, the parameter $a_0$ specifies the radial position of the center of the spiral arm on the $X$ axis. Finally, note that in this method the estimate of the pitch angle $i$ does not depend on the number of spiral arms $m$.

The pitch angle of the spiral pattern of the Galaxy, $i$, was estimated from the distribution of masers using the relation~(\ref{spiral-03}) in Bobylev and Bajkova (2013; 2014). As a result, for four segments of spiral arms, the estimate $i=-13\pm1^\circ$ was obtained.

\begin{figure}[t]
    \centering
    \includegraphics[width = 0.95\textwidth]{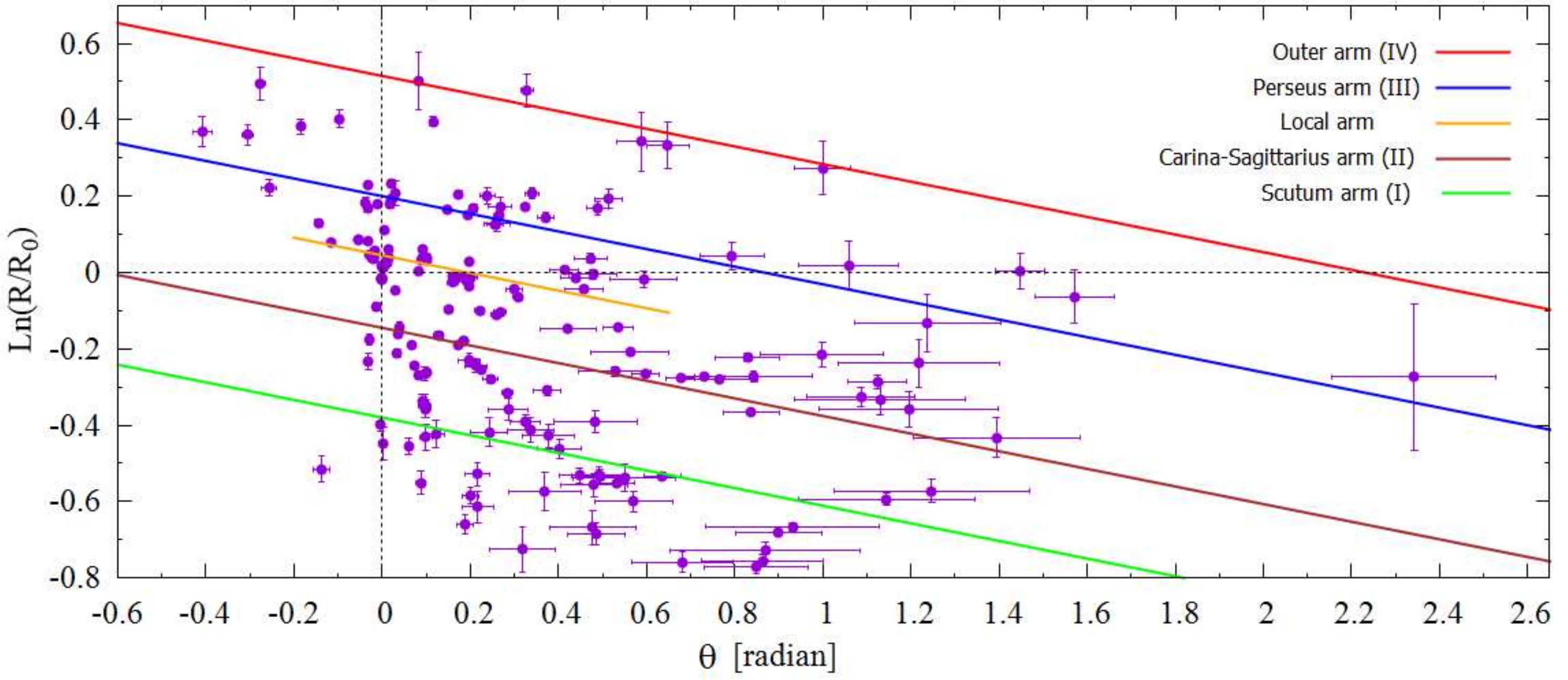}
    \caption{Distribution of masers and radio stars with trigonometric parallax errors less than 15\% on the ``position angle~-- logarithm of distance'' plane, the slope of all lines corresponds to the pitch angle $i=-13^\circ$, the Sun is at the origin.}
    \label{f-15-iii}
\end{figure}

Figure~\ref{f-15-iii} shows the distribution of 204 masers and radio stars on the plane ``position angle~-- logarithm of distance''. This is the same sample that was used to construct Fig.~\ref{f-15-XY}. In Fig.~\ref{f-15-iii} one can distinguish the Outer Arm, the Perseus Arm, the Local Arm and, to a lesser extent, the Carina-Sagittarius Arm. It is impossible to single out the objects of the arm of the Scutum in this figure.

Until now, there is no generally accepted model of the global (grand design) spiral structure of the Galaxy. Models with a different number of spiral arms, with a constant or variable pitch angle, symmetric or asymmetric spirals are discussed. Modern data on the distribution of clouds of neutral and ionized hydrogen, as well as maser sources with trigonometric parallaxes measured by the VLBI method, rather speak of a four-arm model with a constant value of the pitch angle, contained in the interval 10--14$^\circ$. A large evidence base in favor of the four-armed global pattern is collected in reviews by Vall\'ee (1995; 2002; 2008; 2017).

\subsection{Perturbation velocities}
The influence of the spiral density wave in radial $V_R$ (along the radius from the axis of rotation of the Galaxy) and residual tangential velocities $\Delta V_{circ}$ is periodic with an amplitude of about 10--15~km/s. According to the linear theory of density waves (Lin, Shu, 1964), the perturbation velocities satisfy the following relations:
 \begin{equation}
 \begin{array}{lll}
       V_R =-f_R \cos \chi,\\
 \Delta V_{circ}= f_\theta \sin\chi,
 \label{DelVRot}
 \end{array}
 \end{equation}
where
 \begin{equation}
 \chi=m[\cot(i)\ln(R/R_0)-\theta]+\chi_\odot
 \end{equation}
is a phase of the spiral wave ($m$~--- number of spiral arms, $i$~is a pitch angle of the spiral pattern ($i<0$ for a twisting spiral), $\chi_\odot$~is a radial phase of the Sun in a spiral wave); $f_R$ and $f_\theta$ are amplitudes of perturbations of radial and tangential velocities, which are considered as positive values.
The wavelength $\lambda$ (the distance between adjacent segments of the spiral arms, counted along the radial direction) is calculated based on the relation
\begin{equation}
 2\pi R_0/\lambda=m\cot(|i|).
 \label{a-04}
\end{equation}

To study periodic deviations of masers from the rotation curve of the Galaxy, Bajkova and Bobylev (2012) proposed for the first time a method based on periodogram Fourier analysis, which takes into account the logarithmic nature of the spiral structure of the Galaxy, as well as the position angles of galactic objects. The method makes it possible to carry out an accurate analysis of the velocities of objects distributed over a wide range of galactocentric distances.

Bobylev and Bajkova (2013) used this method to analyze a sample of 73 masers distributed over a wide region of the Galaxy, with Galactocentric distances $R:$ 0--20 kpc, the linear rotation velocity of the Galaxy $V_0$ was estimated, and perturbation velocities $f_R$ and $f_\theta$ in the radial and tangential directions, respectively.

Estimates of these parameters are regularly refined by various authors as the sample of masers increases. Bobylev and Bajkova (2022), the last in this series, used this method to analyze a sample of masers and radio stars whose parallaxes were determined with errors of less than 10\%. The value of the found rotation velocity of the Galaxy is indicated in the table~\ref{t:R0}. The amplitudes of the disturbance velocities $f_R=6.7\pm1.1$~km s$^{-1}$ and $f_\theta=2.6\pm1.2$~km s$^{-1}$ were also determined, the wavelength of the disturbance $\lambda=2.1\pm0.3$~ kpc and phase of the Sun in the spiral wave $-148\pm15^\circ$.

Bobylev and Bajkova (2015) showed for the first time the presence of periodic perturbations in the vertical velocities of masers. In particular, the vertical disturbance velocity amplitude $f_W=3.4\pm0.7$~km s$^{-1}$ and the disturbance wavelength $\lambda=4.3\pm1.2$~kpc were found. The presence of vertical perturbations in maser velocities was confirmed, for example, in the works of Bobylev et al. (2016a), Rastorguev et al. (2017). In the work by Bobylev and Bajkova (2022) mentioned above, a new estimate of the vertical velocity perturbation amplitude $f_W=5.2\pm1.5$~km s$^{-1}$ was obtained from masers.

\begin{figure}[t]
    \centering
    \includegraphics[width = 0.85\textwidth]{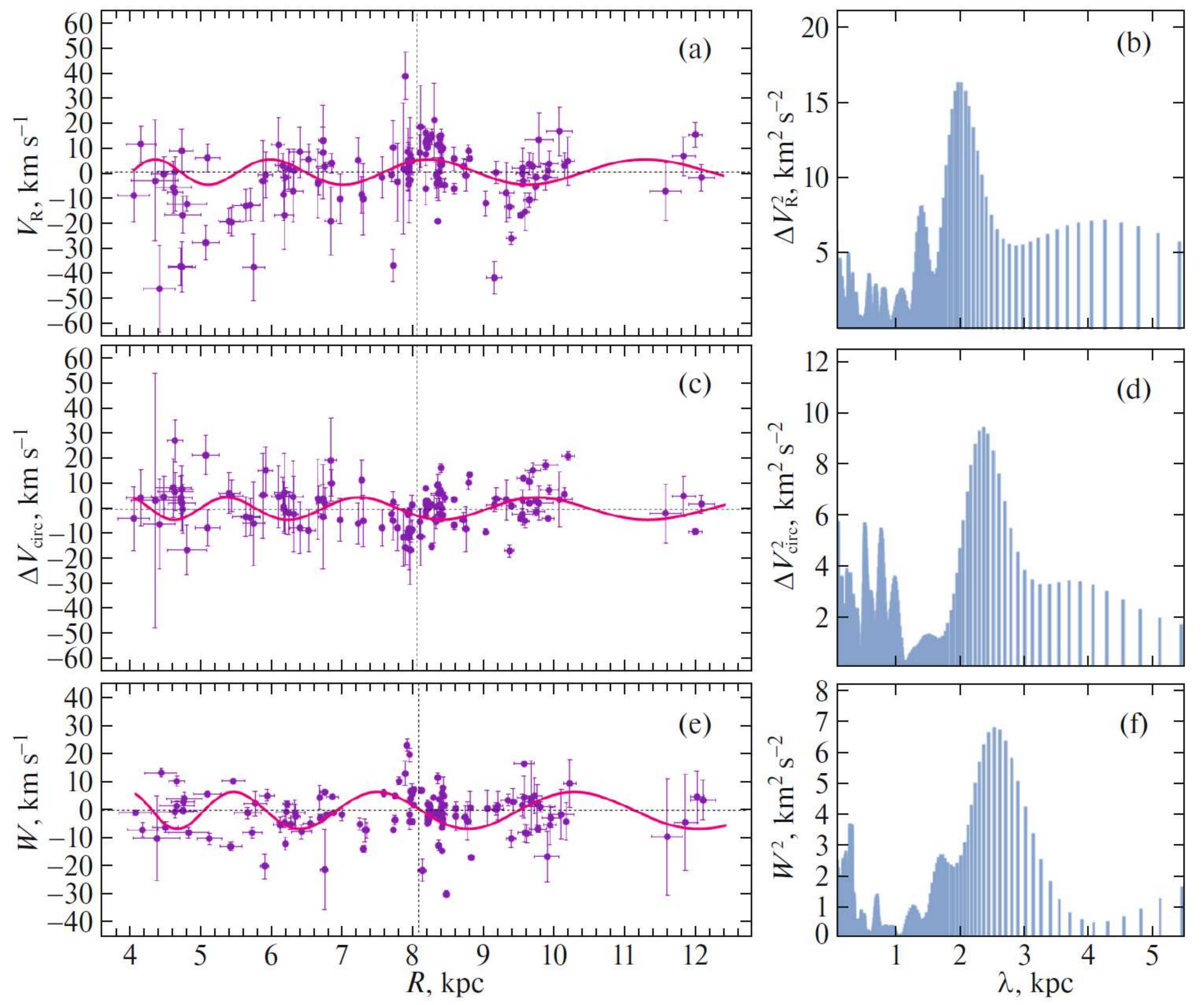}
\caption{Radial velocities $V_R$ of masers versus distance $R$~(a) and their power spectrum~(b), residual rotation velocities  $\Delta V_{circ}$ of masers versus $R$~(c) and their power spectrum~(d), vertical velocities $W$ of masers versus $R$~(e) and their power spectrum~(f), solid wavy lines reflect the results of spectral analysis. Here we use 134 maser sources with relative trigonometric parallax errors less than 10\%, located no farther than 5~kpc from the Sun. The figure is taken from Bobylev and Bajkova (2022).}
    \label{f-spectr}
\end{figure}

Figure~\ref{f-spectr} shows the results of a spectral analysis of a sample of masers and radio stars with relative trigonometric parallax errors less than 10\%, which was performed by Bobylev and Bajkova (2022).

 \section{Other tasks}
 \subsection{Gaia Parallax Zero-Point Offset}
It is known that the trigonometric parallaxes of Gaia (Prusti et al., 2016) have a small shift relative to immobile extragalactic sources (quasars). A small shift, difficult to account for, was preserved even in the Gaia\,DR3 version (Vallenari et al., 2022).

For the first time in the work of Lindegren et al. (2018) noted the presence of a systematic offset $\Delta\pi=-0.029$~mas in the Gaia\,DR2 parallaxes (Brown et al., 2018) with respect to the inertial coordinate system. Later, the presence of such a correction was confirmed by many authors on a variety of material, and with very good accuracy.
The correction $|\Delta\pi|$ must be added to the parallaxes of the stars in the Gaia catalog. In this case, the distances to stars calculated from the corrected parallaxes will decrease. Thus, after taking into account the correction, the stars become closer to the Sun.

Thus, based on 89 separated eclipsing binaries, Stassun and Torres (2018) found a correction $\Delta\pi=-0.082\pm0.033$~mas. According to these authors, the relative parallax errors of the eclipsing binaries used do not exceed 5\% on average and do not depend on the distance.
Riess et al. (2018) obtained an estimate of $\Delta\pi=-0.046\pm0.013$~mas from a sample of 50 long-period Cepheids. They used the photometric characteristics of these Cepheids, measured from the space telescope Hubble. In the work of Zinn et al. (2019) from a comparison of the distances of about 3000 giants from the APOKAS-2 catalog, the value $\Delta\pi=-0.053\pm0.002$~mas was found.

Bobylev (2019) estimated the link parameters between the optical and radio systems using the data from the Gaia\,DR2 catalog and VLBI measurements using a sample of 88 masers and radio stars. For this purpose, both masers and various radio stars (both young and evolved giants of the asymmetric branch) observed in the continuum were used. A new estimate of the systematic offset between the optical and radio systems was obtained: $\Delta\pi=-0.038\pm0.046$~mas as a weighted average of the parallax differences of radio stars like ``Gaia--VLBI''. As you can see, the result is obtained with a large uncertainty.
Xu et al. (2019) repeated Bobylev's analysis for almost the same sample of stars. However, they separately studied the differences between young and old, single and multiple stars. As a result, based on a sample of 34 young single radio stars, these authors found $\Delta\pi=-0.075\pm0.029$~mas. Even earlier, from a comparison of the parallaxes of 55 radio stars in the Gould Belt, Kounkel et al. (2018) determined $\Delta\pi=-0.073\pm0.034$~mas.

\begin{figure}[t]
    \centering
    \includegraphics[width=0.75\textwidth]{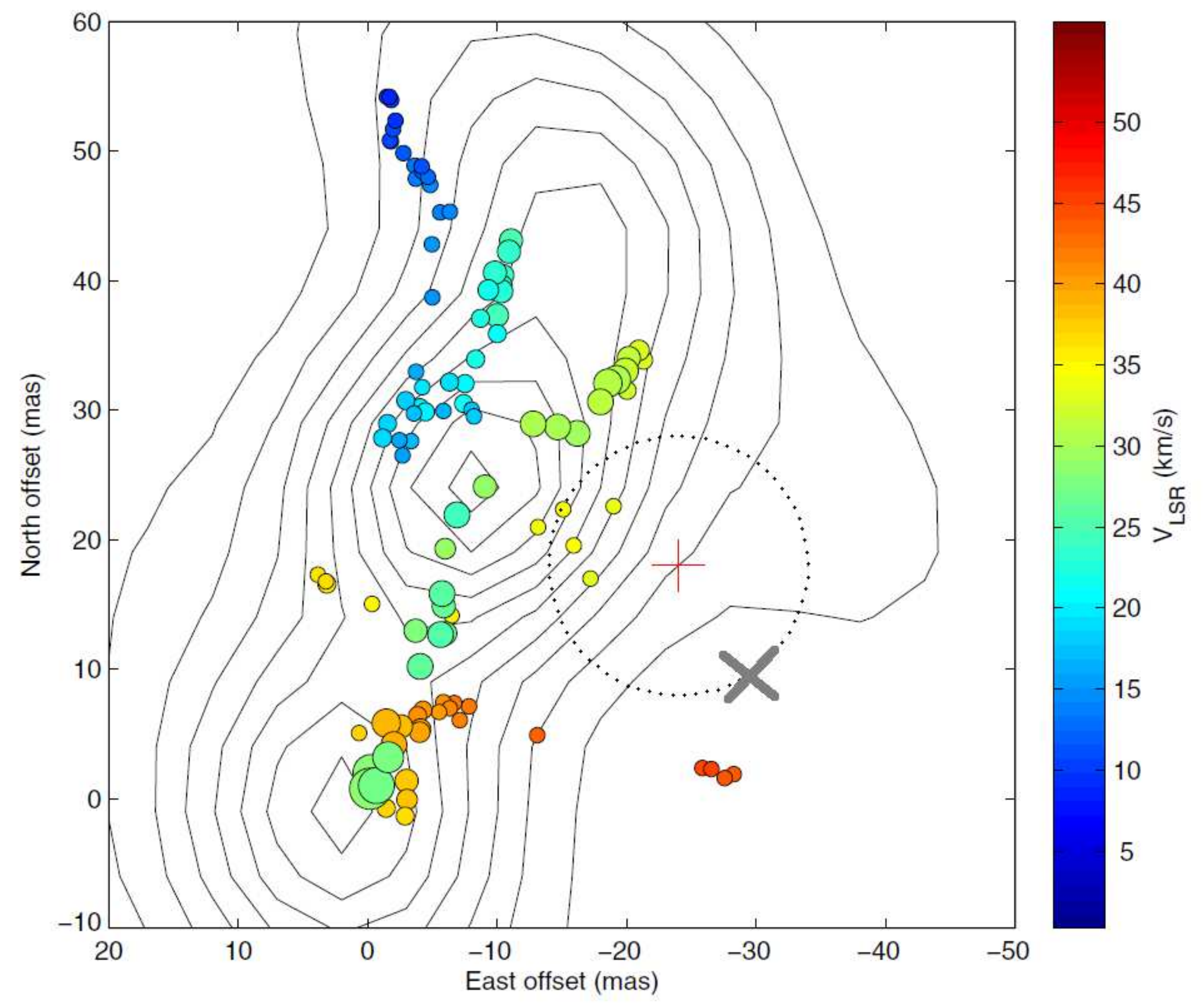}
    \caption{
Distribution of maser spots around the supergiant VY\,CMa. The red cross marks the radio position of the center of the star, the circle of dotted lines indicates the boundary of its atmosphere with a size of 10~mas, and the center of the optical image of the star is shown with an oblique cross. The Figure is taken from Zhang et al. (2012).
}
    \label{f-VY-CMa}
\end{figure}

The relatively large errors in the estimates of the link parameters between the optical and radio systems show that there is also a problem of displacement between the image centers of the optical and radio images. In optical observations of single stars, the photocenter usually lies very close to the direction towards the physical center of the star. Masers, on the other hand, are often distributed asymmetrically in the wide gas-dust envelope of the star.

An example of a discrepancy between optical and radio images is the supergiant VY\,CMa (Zhang et al., 2012), whose radio map is given in Fig.~\ref{f-VY-CMa}. It can be seen that the maser spots are located asymmetrically in a very wide region around the star, and this figure shows only a part of the maser spots. For it, the parallax difference between the radio and the measurement in the Gaia DR2 catalog was $-6.772\pm0.827$~mas (Bobylev, 2019). However, when compared with the Gaia\,EDR3 catalogue, the parallax differences significantly decreased (to values less than 2.5~mas), moreover, due to the improvement of optical measurements.

An example of a symmetrical radio image is the supergiant S\,Per, the results of radio observations of which are given in Fig.~\ref{f-S-Per}. For this star, the parallax difference between radio and optical measurements was $-0.191\pm0.123$~mas (Bobylev, 2019).

\begin{figure}[t]
    \centering
    \includegraphics[width=0.95\textwidth]{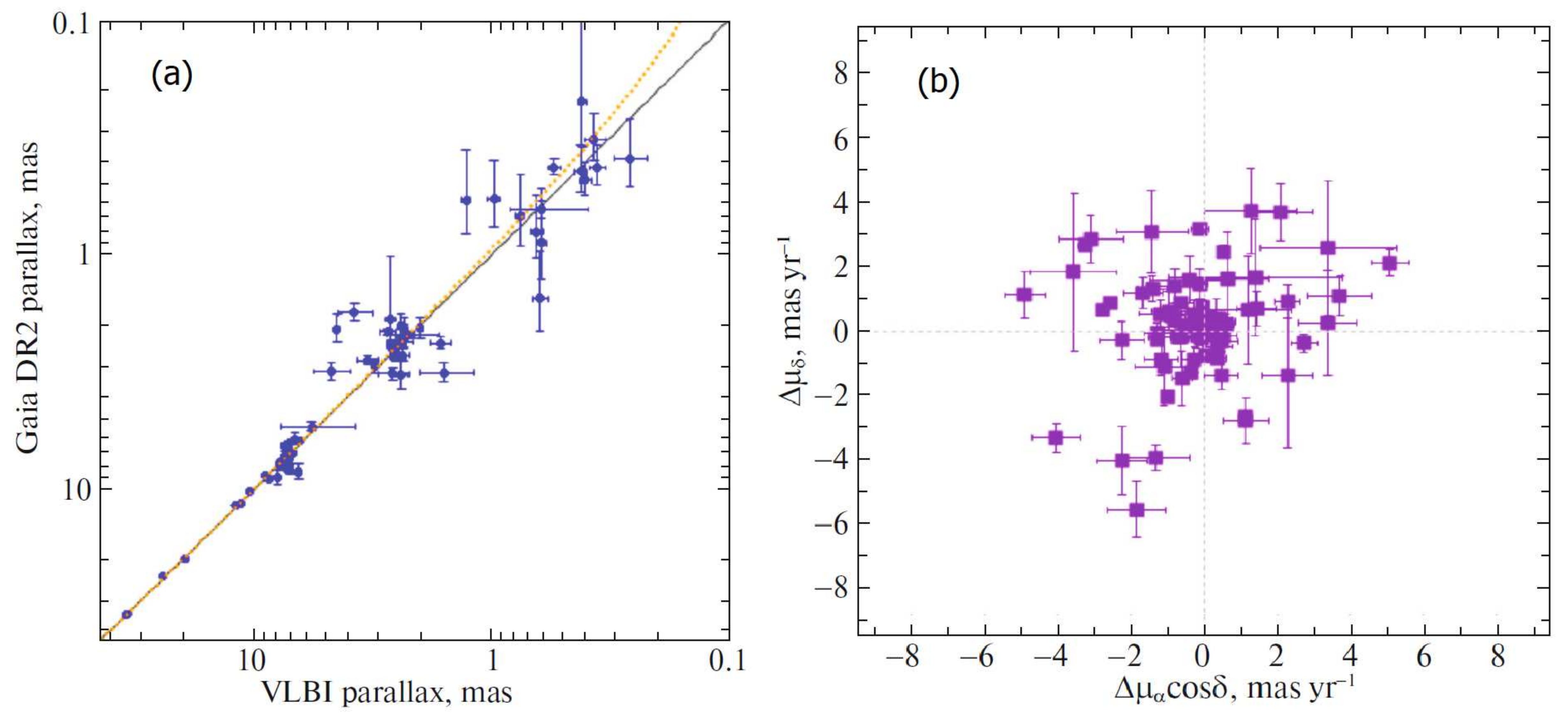}
    \caption{
Parallaxes of radio stars from the Gaia\,DR2 catalog versus their VLBI parallaxes (the solid line corresponds to the correlation with a coefficient of 1) (a), the difference in proper motions of ``Gaia--VLBI'' stars (b). The Figure is taken from Bobylev (2019).
}
    \label{f-diff}
\end{figure}

To compare parallaxes, Bobylev (2019) also used a system of conditional equations of the form:
 \begin{equation}
 \begin{array}{lll}
  \pi_{\rm (Gaia)}=a+b\cdot\pi_{\rm (VLBI)},
 \label{EQ-111}
 \end{array}
\end{equation}
from the solution of which estimates of the parameters $a$ and $b$ were obtained by the method of least squares. As a result, it was shown that the scale factor $b$ is close to unity, $b=1.002\pm0.007$. Figure~\ref{f-diff}(a) shows the parallaxes of radio stars from the Gaia\,DR2 catalog versus their VLBI parallaxes, where the red dotted line corresponds to the found solution.

Systematic errors present in astrometry and broadband photometry in the Gaia\,EDR3 catalog have been transferred to the latest version of the catalog --- in Gaia\,DR3. In particular, analysis of the Gaia\,EDR3 data allowed us to conclude that the Gaia\,DR3 parallax offset relative to quasars is $\sim-0.017$~mas (Lindegren et al., 2021).

\subsection{Mutual rotation of optical and radio systems}
To determine the rates of mutual rotation of two systems around the equatorial coordinate axes $\omega_x,\omega_y,\omega_z$, the following constraint equations are usually used:
 \begin{equation}
 \begin{array}{lll}
 \Delta\mu_\alpha\cos\delta=
 -\omega_x\cos\alpha\sin\delta-\omega_y\sin\alpha\sin\delta+\omega_z\cos\delta,\\
 \Delta\mu_\delta= +\omega_x\sin\alpha-\omega_y\cos\alpha,
 \label{DR2-VLBI}
 \end{array}
 \end{equation}
where the differences in the proper motions of stars of the form ``Gaia-VLBI'' are found on the left-hand sides of the equations. Bobylev (2019) used 81 differences in the proper motions of stars, the moduli of which do not exceed 6 mas yr$^{-1}$. The distribution of these differences is given in Fig.~\ref{f-diff}(b). The following values were found from them:
 \begin{equation}
 \begin{array}{lll}
 (\omega_x,\omega_y,\omega_z)=(-0.14,0.03,-0.33)\pm(0.15,0.22,0.16)~\hbox{mas yr$^{-1}$},
 \label{w1w2w3-p2}
 \end{array}
 \end{equation}
on the basis of which it was concluded that there are no velocities significantly different from zero mutual rotation.

Lindegren (2020) compiled a sample of 41 bright stars with proper motions from the Gaia\,DR2 catalog, which also have VLBI measurements. As a result, he used 26 differences in the proper motions of stars, from which the following estimates were obtained:
 \begin{equation}
 \begin{array}{lll}
 (\omega_x,\omega_y,\omega_z)=(-0.07,-0.05,-0.01)\pm(0.05,0.05,0.07)~\hbox{mas yr$^{-1}$}.
 \label{w1w2w3-Lindegren}
 \end{array}
 \end{equation}
Here, the conclusion about the absence of mutual rotation velocities significantly different from zero is even more obvious.

 \subsection{Mass of the Galaxy}
Masers with measured trigonometric parallaxes give a reliable estimate of the value of the local rotation velocity of the Galaxy $V_0$ at a near-solar distance. Therefore, at present, these data are widely used to construct a galactic rotation curve. Modern rotation curves of the Galaxy are constructed in a very wide range of distances, up to the outer boundary of the Galaxy ($R\sim300$~kpc). Of course, kinematic data on various objects, Cepheids, globular clusters, dwarf satellite galaxies of the Milky Way, etc., are used to construct such a curve (Sofue, 2012; Bhattacharjee et al., 2014).

Bajkova and Bobylev (2017) used data on masers to refine the gravitational potential of the Galaxy for six potential models. All six are axisymmetric three-component models consisting of a bulge, a disk, and a dark matter halo. It was concluded that the best of those considered is model~III, in which the dark matter halo is described by the NFW relation (Navarro, Frenk, White, 1997). As a result, the mass of the Galaxy enclosed inside a sphere with a radius of 200~kpc was found to be $M_{200}=(0.75\pm0.19)\times10^{12}M_\odot.$

Ten years ago, estimates of the mass of the Galaxy differed by orders of magnitude. At present, due to the appearance of mass catalogs with high-accuracy photometric and kinematic data, the scatter of such estimates has been significantly reduced. The mass of the Galaxy enclosed inside a sphere with a radius of 200~kpc, according to various estimates, lies in the range $(0.5-1.5)\times10^{12}M_\odot$ (see Fig.~14 in Wang et al. (2022) ).

We note the latest publications on this topic. Bird et al. (2022) found $M_{200}=(1.00^{+0.67}_{-0.33})\times10^{12}M_\odot,$ using blue horizontal branch stars with data from the LAMOST, SDSS/SEGUE, and Gaia catalogs and for K-giants received an estimate
$M_{200}=(0.55^{+0.15}_{-0.11})\times10^{12}M_\odot.$ Wang et al. (2022), based on globular clusters with proper motions from the Gaia\,EDR3 catalog (Brown et al., 2021), it was concluded that the total mass of the Galaxy lies in the interval $(0.54^{+0.81}_{-0.68}- 0.78^{+3.08}_{-1.97})\times10^{12}M_\odot.$

\subsection{Corrections for the Lutz-Kelker effect}
Corrections for the Lutz-Kelker effect (Lutz, Kelker, 1973) play an important role when using the trigonometric parallaxes $\pi$ of stars in estimating the heliocentric distances $r$ based on them, $r=1/\pi$. As shown in the work of Lutz, Kelker (1973), corrections become necessary when the relative errors in the determination of parallaxes are more than 10--15\%.

For a sample of 54 masers with VLBI parallaxes, Stepanishchev and Bobylev (2013) obtained estimates for the corrections for the Lutz-Kelker effect for the first time (for the first time for masers). The largest correction was found for the source IRAS~16293$-$2422, although this source turned out to be one of the closest to the Sun. Therefore, the introduction of the found correction, $-1.175$~mas, did not have a critical effect on the determination, for example, of the distance-dependent parameters of galactic rotation. Corrections for the Lutz-Kelker effect were also taken into account by Rastorguev et al. (2017) in a kinematic analysis of a sample of 131 masers.

With relative errors in determining parallaxes less than 10\%, corrections for the Lutz-Kelker effect can be neglected. Therefore, we consider the results of determining the parameters of the rotation of the Galaxy that we obtained (Bobylev, Bajkova, 2022) to be very reliable under restrictions on the relative errors of the maser parallaxes (see Table~\ref{t:R0}).

\section{The Local arm and the Radcliffe wave}
Near the Sun, the Radcliffe wave is known to propagate along the Local Arm. It was first discovered from an analysis of the distribution of molecular clouds (Alves et al., 2020). The original authors of this research team are from the Radcliffe Institute for Advanced Study in Cambridge, Massachusetts. Therefore, they named the wave in honor of their native institute.

The Radcliffe wave is a narrow chain of molecular clouds tilted about $-30^\circ$ to the galactic axis $Y$. Its main property is that it manifests itself in the vertical coordinates of clouds $Z$. According to Alves et al. (2020) the wave is damped, has a wavelength of about 2~kpc, with a maximum amplitude of about 160~pc. Moreover, the maximum value of the amplitude is observed near the Sun, in the region of the Gould Belt.
The geometrical characteristics of the Radcliffe wave are confirmed by observations of interstellar dust, T\,Tauri, and OB stars. But the kinematic and dynamic properties of this wave are still poorly understood.

%%%%%%%%%%%%%%%%%%%%%%%%%%%%%%%%%%%%%%%%%%%%%% FIG. WR:
\begin{figure}[t]{ \begin{center}
  \includegraphics[width=0.95\textwidth]{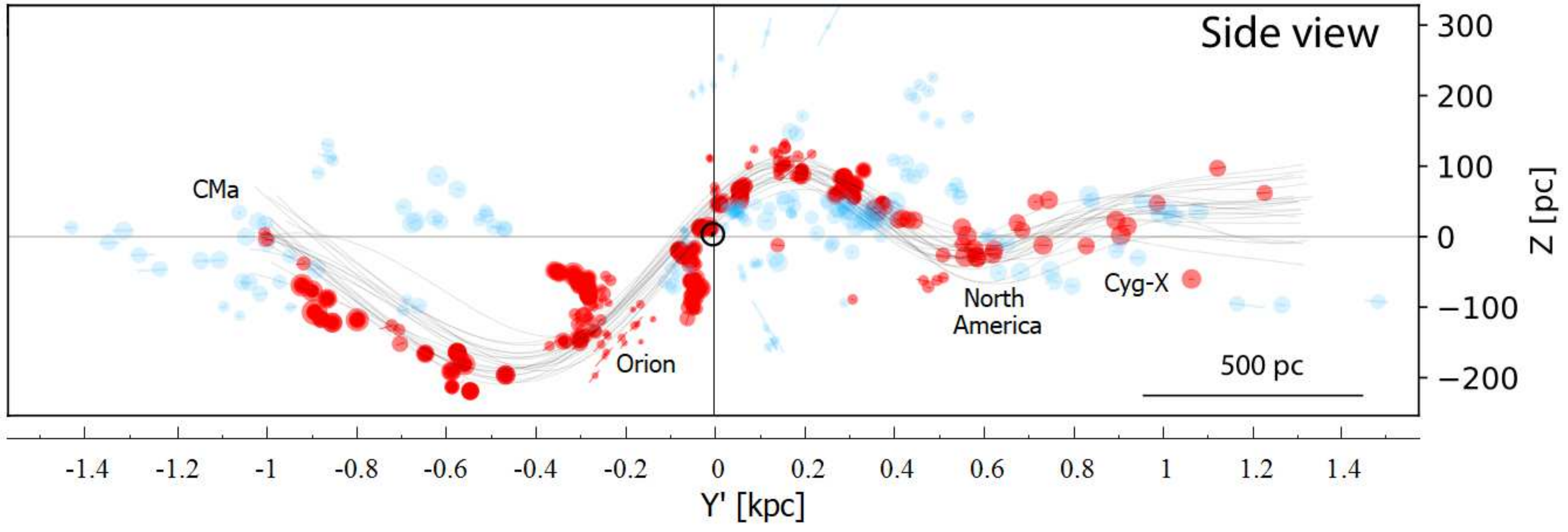}
  \caption{
Vertical coordinates of molecular clouds $Z$ versus the position on the $Y'$ axis (this axis is at an angle of $-30^\circ$ to the galactic axis $Y$), the red circles indicate the clouds tracing the Radcliffe wave, the blue circles indicate field clouds, gray dotted lines are Radcliffe wave models. The Figure is taken from Alves et al. (2020), to which we have added a more detailed $Y'$ scale.}
 \label{f-Alves-00}\end{center}}\end{figure}
%%%%%%%%%%%%%%%%%%%
%%%%%%%%%%%%%%%%%%%%%%%% FIG.6:
\begin{figure}[t]
{ \begin{center}
  \includegraphics[width=0.95\textwidth]{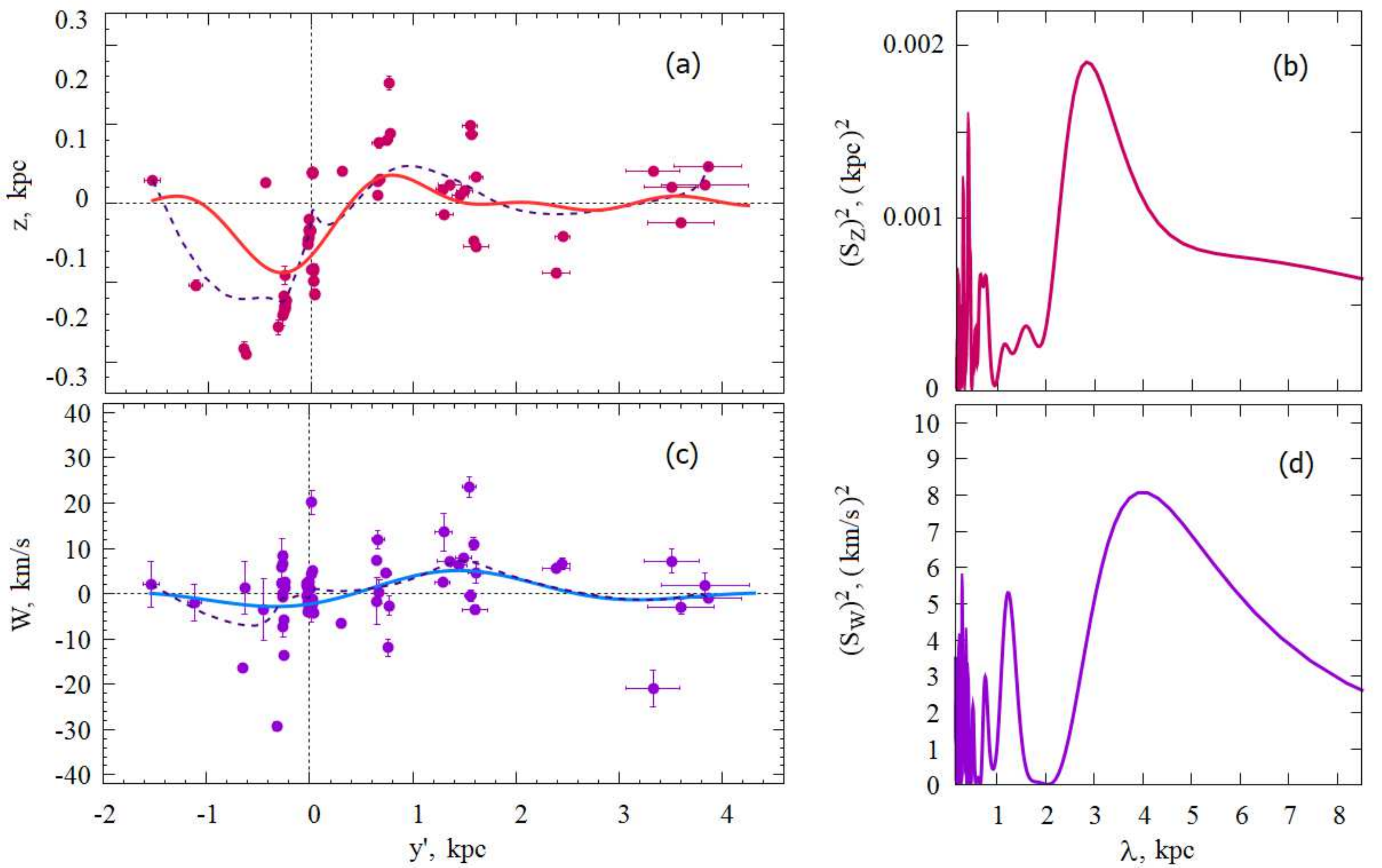}
  \caption{
$Z$ maser coordinates versus distance $Y'$~(a) and their power spectrum~(b), vertical velocities of $W$ masers versus distance $Y'$~(c) and their power spectrum~( d), periodic curves shown by solid thick lines reflect the results of spectral analysis, dotted lines show smoothed average values. The Figure is taken from Bobylev et al.~(2022).
 }
 \label{f-spectr-masers}
\end{center}}
\end{figure}
%%%%%%%%%%%%%%%

Fig.~\ref{f-Alves-00} shows the vertical coordinates of the Alves et al. (2020) molecular clouds. Here they are located along the $Y'$ axis, which is oriented at an angle of $-30^\circ$ to the galactic $Y$ axis. The figure shows a number of the Radcliffe wave models that were found in the work of these authors.

Bobylev, Bajkova (2022), Bobylev et al. (2022) show that masers belonging to the Local Arm demonstrate the presence of the Radcliffe wave. A total of 68 such masers and radio stars were selected for analysis. Based on spectral analysis, the parameters of the Radcliffe wave, such as the wavelength $\lambda=2.8\pm0.1$~kpc, and the amplitude of deviation from the symmetry plane $Z_{max}=87\pm4$~pc, were determined from them. For the first time, the amplitude of the perturbation of vertical velocities $W=5.1\pm0.7$~km/s with a wavelength $\lambda=3.9\pm1.6$~kpc is estimated.

The results of spectral analysis are shown in Fig.~\ref{f-spectr-masers}. The dotted lines in Fig.~\ref{f-spectr-masers}(a) and (c) show the smoothed averages of the data. The good agreement in the behavior of the solid and dotted lines in the circumsolar region indicates the reliability of the performed spectral analysis.

\section*{Conclusion}
Let us single out the most important results obtained by us from the analysis of masers and radio stars with trigonometric parallaxes and proper motions measured by the VLBI method.

1).~As the sample of masers and radio stars increases, the rotation parameters of the Galaxy are regularly refined. We are talking about representatives of the young, most rapidly rotating around the galactic center stellar fraction of the thin disk of the Galaxy. In one of the last works in this series, based on a sample of masers and radio stars with parallax errors less than 10\%, the linear velocity of rotation of the Galaxy at a near-solar distance for the accepted value $R_0=8.1\pm0.1$~kpc was found equal to $V_0=244.4\pm4.3$~km s$^{-1}$.

2).~A review of estimates of the geometric characteristics of the spiral pattern of the Galaxy is given. At present, it is believed that a four-armed spiral pattern with a constant twist angle $i$ with a value of 10--14$^\circ$ is most likely realized in the Galaxy.

3).~The results of determining the perturbation velocities $f_R$ and $f_\theta$ caused by the influence of the galactic spiral density wave on the motion of young stars are presented. It is important to note that such perturbations are also found in the vertical velocities of masers, $f_W$.

4).~A review of estimates of the parameters of the vertical distribution of stars in the thin disk of the Galaxy is given. In particular, the value of the vertical disk scale $h$ under the assumption of an exponential distribution of matter, found by us from methanol masers, is $26.5\pm0.7$~pc.

5).~A short review of works devoted to the refinement of the gravitational potential and estimates of the mass of the Galaxy is made. According to modern estimates, the mass of the Galaxy enclosed inside a sphere with a radius of 200~kpc lies in the range $(0.5-1.5)\times10^{12}M_\odot$.

6).~Works devoted to masers and radio stars belonging to the Local Arm are noted. There is evidence that they demonstrate the presence of a Radcliffe wave. The wavelength and amplitude of deviation from the symmetry plane found from them are in good agreement with those obtained from molecular clouds in the pioneering work of Alves et al. (2020). And in the work of Bobylev et al. (2022), the velocity amplitude of vertical disturbances in the Radcliffe wave, $W=5.1\pm0.7$~km s$^{-1}$, with a wavelength $\lambda=3.9\pm1.6$~kpc, was determined from masers for the first time.

7). The use of masers and radio stars with measured VLBI parallaxes for estimating the zero-point shift of the Gaia\,DR2 parallaxes and determining the parameters of the mutual rotation of the radio and optical systems is described.

  \medskip
The authors are grateful to the referee for useful comments that contributed to the improvement of the manuscript.

\medskip

{REFERENCES}\medskip {\small\begin{enumerate}

 \item
Abuter et al. (GRAVITY Collaboration, R. Abuter, A. Amorim, N. Baub\"ock,
  et al.), Astron. Astrophys. {\bf 625}, L10 (2019).

 \item
Abuter et al. (GRAVITY Collaboration, R. Abuter, A. Amorim, M. Baub\"ock,
 et al.), Astron. Astrophys. {\bf 647}, A59 (2021).

 \item
J. Alves, C. Zucker, A.A. Goodman, et al., Nature {\bf 578}, 237 (2020).

 \item
Y. Asaki, S. Deguchi, H. Imai, et al, Astrophys. J. {\bf 721}, 267 (2010).

  \item
A.T. Bajkova, V.V. Bobylev, Astron. Lett. {\bf 38}, 549 (2012).

 \item
A.T. Bajkova, V.V. Bobylev, Baltic Astron. {\bf 24}, 43 (2015).

 \item
A.T. Bajkova, V.V. Bobylev, Open Astron. {\bf 26}, 72 (2017).

 \item
S.A. Bird, X.-X. Xue, C. Liu, et al., MNRAS {\bf 516}, 731 (2022).

 \item
V.V. Bobylev, A.T. Bajkova, MNRAS {\bf 408}, 1788 (2010).

 \item
V.V. Bobylev, A.T. Bajkova, Astron. Lett. {\bf 39}, 759 (2013).

 \item
V.V. Bobylev, A.T. Bajkova, MNRAS {\bf 437}, 1549 (2014).

 \item
V.V. Bobylev, A.T. Bajkova, MNRAS {\bf 447}, L50 (2015).

 \item
V.V. Bobylev, A.T. Bajkova, and  K.S. Shirokova, Baltic Astron. {\bf 25}, 15 (2016).

 \item
V.V. Bobylev, A.T. Bajkova, Astron. Lett. {\bf 42}, 182 (2016a).

 \item
V.V. Bobylev, A.T. Bajkova, Astron. Lett. {\bf 42}, 210 (2016b).

 \item
V.V. Bobylev, Astron. Lett. {\bf 45}, 10 (2019).

 \item
V.V. Bobylev, O.I. Krisanova, and  A.T. Bajkova, Astron. Lett. {\bf 46}, 439 (2020).

 \item
V.V. Bobylev, A.T. Bajkova, Astron. Rep. {\bf 65}, 498 (2021).

 \item
V.V. Bobylev, A.T. Bajkova, Astron. Lett. {\bf 48}, 376 (2022).

 \item
V.V. Bobylev, A.T. Bajkova, and Yu.N. Mishurov, Astron. Lett. {\bf 48},(2022).

 \item
P. Bhattacharjee, S. Chaudhury, and S. Kundu, Astrophys. J. {\bf 785}, 63 (2014).

 \item
Brown et al.  (Gaia Collaboration, A.G.A. Brown, A. Vallenari, T. Prusti,
 et al.), Astron. Astrophys. {\bf 616}, 1 (2018).

 \item
Brown et al. (Gaia Collaboration, A.G.A. Brown, A. Vallenari, T. Prusti,
 et al.), Astron. Astrophys. {\bf 649}, 1 (2021).

 \item
F. Gao, J.A. Braatz, M.J. Reid, et al., Astrophys. J. {\bf 834}, 52 (2017).

 \item
K. Hachisuka, A. Brunthaler, K.M. Menten, et al., ApJ  {\bf 645}, 337 (2006).

 \item
M. Honma, T. Nagayama, K. Ando, et al., PASJ {\bf 64}, 136 (2012).

 \item
Hirota et al. (VERA collaboration, T. Hirota, T. Nagayama, M. Honma,
et al.), PASJ {\bf 70}, 51 (2020).

 \item
K. Immer, K.L.J. Rygl, Universe {\bf 8}, 390 (2022).

\item
V. Krishnan,  S.P. Ellingsen,  M.J. Reid, et al., Astrophys. J. {\bf 805}, 129 (2015).

\item
M. Kounkel, K. Covey, G. Suarez, et al., Astron. J. {\bf 156}, 84 (2018).

\item
Lindegren et al. (Gaia Collaboration, L. Lindegren, J. Hernandez, A. Bombrun, et al.),
  Astron. Astrophys. {\bf 616}, 2 (2018).

\item
L. Lindegren, Astron. Astrophys. {\bf 637}, C5 (2020).

\item
Lindegren et al. (Gaia Collaboration, L. Lindegren, U. Bastian, M. Biermann, et al.),
  Astron. Astrophys. {\bf 649}, 4 (2021).

\item
C.C. Lin, F.H. Shu, Astrophys. J. {\bf 140}, 646 (1964).

\item
A.V. Loktin, M.E. Popova, Astrophys. Bull. {\bf 74}, 270 (2019).

\item
T.E. Lutz, D.H. Kelker, PASP {\bf 85}, 573 (1973).

\item
J. Maiz-Apell\'aniz, Astron. J. {\bf 121}, 2737 (2001).

\item
D.J. Marshall, A.C. Robin, C. Reyl\'e, et al. Astron. Astrophys. {\bf 453}, 635 (2006).

\item
J.C.A. Miller-Jones, P.G. Jonker, V. Dhawan, et al., Astrophys. J. {\bf 706}, L230 (2009).

\item
L. Moscadelli, A. Sanna, H. Beuther, arXiv: 2209.00432 (2022).

\item
T. Nagayama, T. Hirota, M. Honma, et al., PASJ {\bf 72}, 51 (2020).

\item
J.F. Navarro, C.S. Frenk, and S.D.M. White, Astrophys. J. {\bf 490}, 493 (1997).

 \item
G.N. Ortiz-Le\'on, L. Loinard,  M.A. Kounkel, et al., Astrophys. J. {\bf 834}, 141 (2017).

 \item
M. Persic, P. Salucci, and F. Stel, MNRAS {\bf 281}, 27 (1996).

 \item
Prusti et al. (Gaia Collaboration,  T. Prusti, J.H.J. de Bruijne, A.G.A. Brown,
   et al.), Astron. Astrophys. {\bf 595}, 1 (2016).

 \item
A.S. Rastorguev, N.D. Utkin, M.V. Zabolotskikh, et al.,
Astrophys. Bull. {\bf 72}, 122 (2017).

 \item
A.G. Riess, S. Casertano, W. Yuan, et al., Astrophys. J. {\bf 861}, 126 (2018).

 \item
M.J. Reid, K.M. Menten, X.W. Zheng, et al., Astrophys. J. {\bf 700}, 137 (2009).

 \item
M.J. Reid, J.E. McClintock, J.F. Steiner, et al., Astrophys. J. {\bf 796}, 2 (2014).

 \item
M.J. Reid, K.M. Menten, A. Brunthaler, et al., Astrophys. J. {\bf 885}, 131 (2019).

 \item
N. Sakai, B. Zhang, S. Xuet, et al., arXiv: 2211.12534 (2022).

 \item
D.M. Skowron, J. Skowron, P. Mr\'oz, et al., Science {\bf 365}, 478 (2019).

 \item
Y. Sofue, PASJ {\bf 64}, 75 (2012).

 \item
L. Spitzer, Astrophys. J. {\bf 95}, 329 (1942).

 \item
K.G. Stassun, G. Torres, Astrophys. J. {\bf 862}, 61 (2018).

 \item
A.S. Stepanishchev,  V.V. Bobylev, Astron. Lett. {\bf 37}, 254 (2011).

 \item
A.S. Stepanishchev,  V.V. Bobylev, Astron. Lett. {\bf 39}, 185 (2013).

 \item
Vallenari et al. (Gaia Collaboration, A. Vallenari, A.G.A. Brown, T. Prusti, et al.),
arXiv: 2208.00211 (2022).

 \item
J.P. Vall\'ee, Astrophys. J. {\bf 454}, 119 (1995).

 \item
J.P. Vall\'ee, Astrophys. J. {\bf 566}, 261 (2002).

 \item
J.P. Vall\'ee, Astron. J. {\bf 135}, 1301 (2008).

 \item
J.P. Vall\'ee, New Astron. Review {\bf 79}, 49 (2017).

 \item
J. Wang, F. Hammer, and Y. Yang, MNRAS {\bf 510}, 2242 (2022).

 \item
Y. Xu, M.J. Reid, X.W. Zheng, and K.M. Menten, Science {\bf 311}, 54 (2006).

 \item
S. Xu, B. Zhang, M.J. Reid, et al., Astrophys. J. {\bf 875}, 114 (2019).

 \item
B. Zhang, M.J. Reid, K.M. Menten, et al., Astrophys. J. {\bf 744}, 23 (2012).

 \item
J.C. Zinn, M.H. Pinsonneault, D. Huber, and D. Stello, Astrophys. J. {\bf 878}, 136 (2019).

 \end{enumerate}
 }

 \end{document}